\documentclass[useAMS,usenatbib]{mn2e}
\usepackage{graphicx}
   \title{Discovery of a new PG1159 (GW Vir) Pulsator}

   \author[Kepler et al.]{S. O. Kepler$^{1}$\thanks{
kepler@if.ufrgs.br},
Luciano Fraga$^2$,
Don Earl Winget$^3$,
Keaton Bell$^3$,
\newauthor
Alejandro H. C\'orsico$^4$
and
Klaus Werner$^{5}$\\
$^{1}$Instituto de F\'{\i}sica, Universidade Federal do Rio Grande do Sul,
              91501-900  Porto-Alegre, RS, Brazil\\
$^2$Laborat\'orio Nacional de Astrof\'{\i}sica, Itajub\' a, MG, Brazil\\
$^3$Department of Astronomy and McDonald Observatory, The University of Texas at Austin,
Austin TX 78712-1083, USA,\\
$^{4}$Facultad de Ciencias Astron\'omicas y Geof\'{\i}sicas, Paseo del Bosque S/N, 
(1900) La Plata, and\\
   Consejo Nacional de Investigaciones Cient\'{\i}ficas y T\'enicas (CONICET), 
Argentina\\
$^5$Institute for Astronomy and Astrophysics, Kepler Center for Astro and Particle Physics,
Eberhard Karls University, Samd I,\\
72076 T\"ubingen, Germany, e-mail: werner@astr.ini-tuebingen.de
}
\begin{document}
\date{Accepted 2014 May 20.  Received 2014 May 19; in original form 2014 April 9}

\pagerange{\pageref{firstpage}--\pageref{lastpage}} \pubyear{2014}
   \maketitle

\label{firstpage}
  \begin{abstract}

We  report the discovery  of pulsations  in the  spectroscopic PG~1159
type pre-white  dwarf SDSS~J075415.12+085232.18.
Analysis 
of    the    spectrum by \citet{Werner14}    indicated
$T_\mathrm{eff}=120\,000\pm  10\,000$~K, $\log  g=7.0  \pm 0.3$,  mass
${\cal{M}}=0.52 \pm 0.02 \cal{M}_\odot$, C/He=0.33 by number.

We obtained time-series images with the SOAR 4.1~m telescope and 2.1~m
Otto Struve  telescope at  McDonald Observatory and  show the  star is
also a variable PG~1159 type star,
with dominant period of 525~s.     

\end{abstract}


\section{Introduction}

White dwarf stars  are the end product of evolution  of all stars with
initial   masses  up  to   around  $8-10~{\cal{M}}_\odot$,
depending on the metalicity of the progenitor and its effect on mass loss 
and the real
value of the C($\alpha,\gamma$)O reaction rate.  Their spatial and mass
distributions  contain information  about star  formation  history and
subsequent evolution  in our Galaxy.  As the most common  endpoints of
stellar evolution,  white dwarf stars  account for around 95\%  of all
evolved stars.   The GW Vir stars, also called DOVs,
are the pulsating  variables in the
spectroscopic  PG~1159 class  that  links  the (post-AGB)
central  stars of planetary  nebulae and  the H-deficient  white dwarf
cooling sequence.  These stars are  non-radial pulsators and lie in an
instability  strip  bounded  by effective  temperatures  $200\,000\leq
T_\mathrm{eff}  \leq  75\,000$~K,  excited by  the  $\kappa$-mechanism
working   through   partial   ionization   of   carbon   and   oxygen.
Asteroseismological analysis  of these stars  has provided significant
knowledge on  the interiors  of the late  stages of  stellar evolution
\citep{Winget08,Althaus10}.   There  are   20  known   GW   Vir  stars
\cite{Quirion09a,Quirion09b,Woudt12}.   Finding new pulsators  of this  class can
improve  our knowledge  of the  
asymptotic giant branch (AGB) and Very Late Thermal Pulse  (VLT) phases,  as  well as angular
momentum   loss   throughout    the   extensive   mass   loss   phases
\citep{Charpinet09,Corsico11}.

\begin{figure*}    
\begin{minipage}{\textwidth}
\centering    
\includegraphics[height=\textwidth,width=5cm,angle=270]{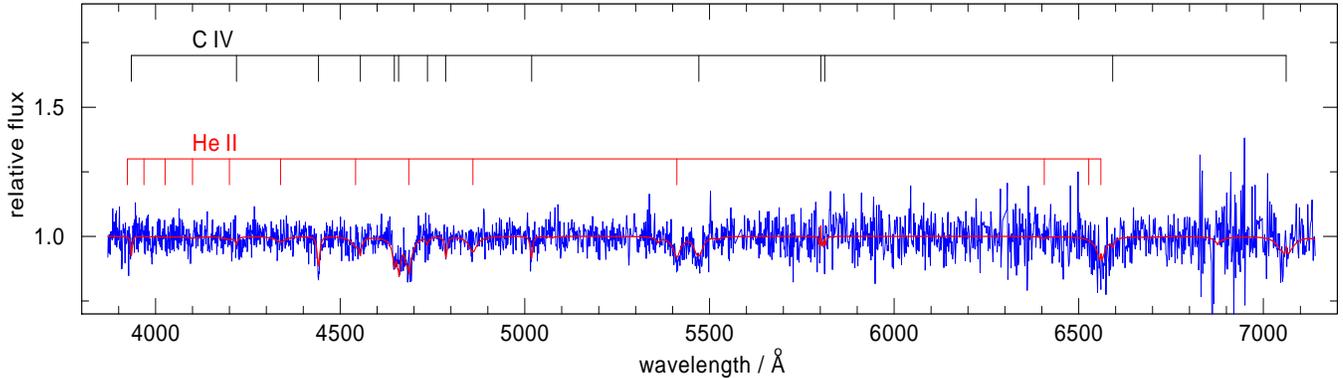}
\caption{Spectrum of SDSS~J075415.12+085232.1 and the model (overplotted in red) with $T_\mathrm{eff}=120\,000$~K,
$\log g=7$ and C/He=0.33 by number.}          
\label{spectra}    
\end{minipage}
\end{figure*}

In our search for new spectroscopically confirmed white dwarf stars in
the  Sloan  Digital  Survey (SDSS)  \citep{Kleinman13},  we  identified
SDSS~J075415.12+085232.18 as  a hot pre-white dwarf  from the presence
of He II and carbon lines in spectrum Plate=2945 MJD=54505 Fiber=183 of
this  g=18.79 star. It shows no detectable planetary nebula.
\citet{Werner14}  fitted  its SDSS spectrum with non local thermodynamic
equilibrium (NLTE)
models  and  obtained  $T_\mathrm{eff}=120\,000\pm  10\,000$~K,  $\log
g=7.0  \pm 0.3$,  mass  ${\cal{M}}=0.52 \pm  0.02 \cal{M}_\odot$,  and
C/He=0.33 by number, indicating the star  is a spectroscopic PG~1159 type star,
i.e., hotter and with a more complex spectrum than a normal DO white dwarf,
similar   to  the   prototype \citep{Liebert89,Werner89},  
which   is  also   a   pulsating  star
\citep[e.g][]{Mcgraw79,Winget91,
Kawaler94,
Costa08a}.
The observed $g=18.79\pm 0.01$ apparent magnitude, compared to an $M_g=5.68$
for such effective temperature and gravity, implies a distance of 
$4.18\pm 0.03$~kpc.
Such a large distance merits the full extinction correction in that direction,
0.076~mag in g, which brings the distance to 4.04~kpc.
\citet{Corsico06a,Corsico06b} computed
fully evolutionary models and non-adiabatic pulsation models  for stars  in the  GW  Vir instability
strip and found that they agree with the observed strip.

\section{Observations \& Data Reduction}

We first obtained time-series photometry of SDSS~J075415.12+085232.18
with  the 4.1-m  SOAR\footnote{Based on  observations obtained  at the
  Southern Astrophysical  Research (SOAR) telescope, which  is a joint
  project of the Minist\'erio  da Ci\^encia, Tecnologia, e Inova\c{c}\~ao (MCTI)
  da  Rep\'ublica  Federativa  do  Brasil,  the  U.S.  National  Optical
  Astronomy Observatory  (NOAO), the  University of  North  Carolina at
  Chapel  Hill  (UNC),  and  Michigan  State  University (MSU).}
telescope and  using the Soar Optical  Imager [SOI, \citet{Schwarz04}]
during  the night of  28 Jan  2014 (Barycentric Julian Terrestrial Time BJTT=245\,6685.7200646). SOI  is a
mini-mosaic  of two E2V  2k$\times$4k CCDs  covering a  5.26~arcmin square
field of  view at a plate  scale of 0.077 arcsec/pixel.  We obtained a
total of 263  SOI frames with a Bessel-B filter,  exposure time of 30~s
and 4$\times$4 binning,  yielding a detector scale of  0.31 arcsec/pixel. The
SOI data frames  were reduced in the standard  manner using the mosaic
reduction (MSCRED) package in IRAF \citep{Valdes98, ValdesTody98}. The
data reduction  process includes bias  subtraction, flat-fielding, and
cosmic-ray  cleaning.  We performed  the  aperture  photometry in  the
individual  frames using {\it  daophot} \citep{Stetson91}  routines in
IRAF. From  the Fourier  analysis, we achieved  a mean noise  level of
$\langle  A\rangle$=1.4~mma,  and  detected   for  the  first  time  a
periodicity,  with a  period of  525~s at  6.8~mma,  therefore at
4.8$\langle A\rangle$,  well above the 1/1000  false alarm probability
limit.
 
On the  three consecutive  nights of  3, 4, and  5 Feb  2014 
(BJTT=245\,6691.699600605) we  obtained follow-up  observations of the  star with
the Cassegrain-mounted ProEM camera  and the PuokoNui data acquisition
software \citep{Chote14}  at McDonald Observatory's  2.1-m Otto Struve
telescope.  From 3127 images with  10 to 30˜s exposures, we confirmed the
525~s  periodicity at  5.9~mma, compared  to the  average  noise level
$\langle A \rangle$=1.03~mma.
The frames were binned at 4$\times$4,
giving a 0.36 arcsec/pixel plate scale across the 2.3$\times$2.3 arcmin field of view.
We performed aperture photometry on the calibrated images using the IRAF package
{\it ccd\_hsp} \citep{Kanaan02} and calculated barycentric corrections 
with the {\it WQED} software \citep{Thompson09}.

We obtained additional time-series  observations with the SOAR Goodman
Spectrograph \citep{Clemens04} in imaging  mode during the night of 27
Feb  2014 (BJTT=245\,6715.5192530).  Goodman is  mounted  at the  SOAR
Optical  Nasmyth and  its  detector is  a  4k$\times$4k Fairchild  486
back-illuminated   CCD,  with   a   un-binned  plate   scale  of   0.15
arcsec/pixel. We carried out the photometric observations with a S8612
red block  filter, a region of interest  (ROI) of 800$\times$800~pixel
square and a 2$\times$2 binning,  yielding a field of view of 4~arcmin
square and a plate scale of  0.3 arcsec/pixel.  Each exposure
lasted 30~s. We used the same method
for  the data  reduction  and photometry  as  for SOI  data. From  the
Fourier  analysis of a  total 451  Goodman frames  we achieved  a mean
noise level of  $\langle A \rangle$=0.94~mma, with which  we were able
to detect three periodicities, 523.5~s at 7.0~mma, 457.2~s at 3.8~mma,
and 439.2 at 3.5~mma, all above the 1/1000 false alarm probability.

\begin{figure}
   \centering
   \includegraphics[width=0.5\textwidth]{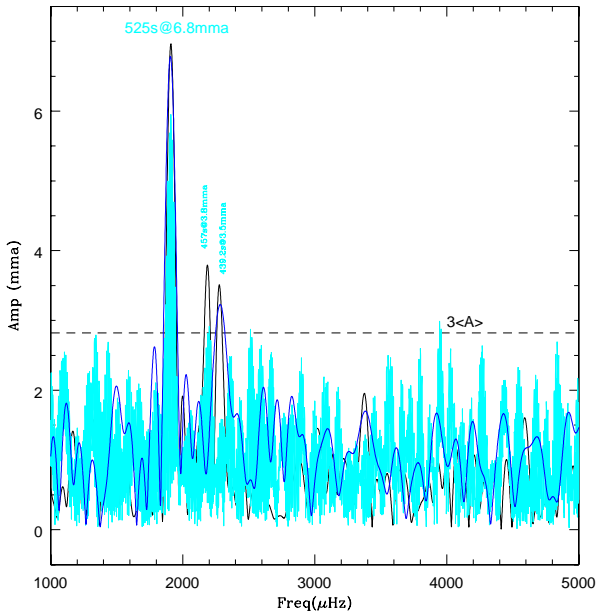}
      \caption{
Fourier transform of the two SOAR data sets (black and blue lines), and the McDonald
data set (cyan shaded). The 3$\langle A \rangle$ line, corresponding to the
false alarm probability of 1/1000, refers only to the 
equally spaced SOAR data set from 27 Feb 2014, the one with lowest noise,
shown in black.
              }
         \label{dft}
   \end{figure}

Figure~\ref{dft} shows the Fourier transform of all data sets.
Analyzing the whole data set at once,
we obtained
523.480$\pm$0.005~s at 5.8$\pm$0.8~mma,
and
524.87$\pm$0.008~s at 3.4$\pm$0.8~mma,
but the different instrument color response 
prevents us from trusting these values.
The spacing in frequency
$\delta \nu\simeq 5.1~\mu$Hz is similar to that
for $\ell=1$ modes of PG~1159$-$035 (4.1$\mu$Hz).
If the spacing is real, it indicates a rotation period
of 28~h, similar to those derived for other variable
PG~1159 stars.

\section{Modelling and seismology}

The pulsation  modelling and seismological analysis  presented in this
section rely  on a set  of stellar models  that take into  account the
complete  evolution  of  PG~1159  progenitor  stars.  The  models  were
extracted from  the evolutionary calculations presented  by \citet{Althaus05}
and \citet{Bertolami06}, who  computed the
complete evolution of model star  sequences with initial masses on the
zero-age main sequence (ZAMS)  ranging from  1  to  $3.75\, M_{\odot}$.   
All  of the  post-AGB
evolutionary   sequences  were   computed  using   the   {\tt  LPCODE}
evolutionary code \citep{Althaus05} and were followed through the
very late  thermal pulse (VLTP)  and the resulting  born-again episode
that gives  rise to the  H-deficient, and He-, C-, and  O-rich composition
characteristic of PG~1159 stars.   The masses of the resulting remnants
are  0.530, 0.542,  0.556, 0.565,  0.589, 0.609,  0.664,  and $0.741\,M_{\odot}$. 

With only three periods detected
for SDSS~J075415.12+085232.18, we  cannot estimate the  mean period
spacing, and cannot constrain the  stellar mass by comparing  with the
mean period spacing of the models,
as  done in the case  of other pulsating  PG~1159 stars 
\citep[e.g.,][]{Corsico09}. 
The way to  infer the stellar mass, along with the
effective temperature  and also details  of the internal  structure of
SDSS~J075415.12+085232.18   is  through  their   individual  pulsation
periods.   This  has  been   the  approach  employed  by
\citet{Corsico07a,Corsico07b, Corsico08, Corsico09} for the  pulsating 
PG~1159 stars
RX~J2117.1+3412,   PG~0122+200,  PG~1159$-$035,  PG~2131+066,
and  
PG~1707+427, respectively.

We  employed the  extensive set  of  $\ell= 1,  2$ $g$-mode  adiabatic
pulsation periods used in \citet{Corsico07a, Corsico07b, Corsico08, Corsico09}.  
For details of  the adiabatic pulsation code ({\tt LP-PUL code}) and methods  employed to
produce  the set  of  periods,  see \citet{Corsico06a}.  We
analyzed more than  about 3000 PG~1159 models covering  a wide range of
effective  temperatures   [$5.4  \ga  \log(T_{\rm   eff})  \ga  4.8$],
luminosities  [$0  \la \log(L_*/  L_{\odot})  \la  4.2$], and  stellar
masses ($0.530 \leq M_*/M_{\odot} \leq 0.741$).
Even though the radial order $k$ associated to the observed periods ($\sim 440-524$ s) is large (as we shall see below), the pulsation $g$-modes of SDSS J075415.12+085232.18 
are probably not in the asymptotic regime (see, for instance, \citet{Corsico06a}).
Because the models are evolutionary, not started from a polytrope,
they cannot achieve any combination of mass, luminosity and effective
temperature, and do not cross each other in the  Hertzsprung-Russell diagram.
The best solutions, quoted, are not just samples of possible solutions,
but limited solutions. As there are three independent modes, one
can estimate up to three parameters of the models.

We  seek  pulsation  models  that  best  match  the
individual   pulsation  periods  of   SDSS~J075415.12+085232.18.   The
goodness  of  the  match  between the  theoretical  pulsation  periods
($\Pi_k$) and  the observed individual periods  ($\Pi_{{\rm obs}, i}$)
is measured by means of a quality function defined as:  
\begin{equation}
\chi^2(M_*, T_{\rm eff})= \frac{1}{N} \sum_{i=1}^{N} 
\min[(\Pi_{{\rm obs},i}- \Pi_k)^2]
\label{ptpf}
\end{equation}
\noindent where $N$ ($= 3$) is the number of observed periods.  In the
absence  of  any additional  information,  we  assume  that the  three
observed periods of SDSS~J075415.12+085232.18 correspond to eigenmodes
with  azimuthal order $m= 0$, but
\citet{Metcalfe03} shows the effect of the assumption is negligible. 
We evaluate  the  function $\chi^2(M_*,T_{\rm eff})$  
for evolutionary models  with stellar masses  of 0.530,
0.542, 0.556,  0.565, 0.589, 0.609, 0.664,  $0.741\,M_{\odot}$. 
The PG~1159 model that  shows the lowest value of $\chi^2$  is adopted as the
``best-fit model''.  Since  we do not know at  the outset the harmonic
degree  ($\ell$) identification  of  the observed  modes,  we have  to
distinguish three cases.

\begin{table}
\centering
\begin{tabular}{cccc}
$\Pi_{\rm obs}$& $\Pi_{\rm theor}$&  $\ell$ & $k$ \cr
439.2&438.9 &1& 18\cr
457.2&458.1 &1& 19\cr
523.5&522.5 &1& 22\cr
\end{tabular}
\caption{Observed and theoretical periods for the best model
fit in Case~1.}
\label{table-case1}
\end{table}

\subsection{Case~(1):  all $\ell= 1$ modes}

Here, we consider  that all the three measured  periods are associated
to modes with  $\ell= 1$. We obtain a  best fit solution characterized
by:   $M_*=  0.556   \,M_\odot$,  $T_{\rm   eff}=  130\,100$   K,  and
$L/L_\odot$= 170.   A comparison between the  observed and theoretical
periods, along with  the derived $\ell$ and $k$  (radial order) values
associated to this solution is shown in Table \ref{table-case1}.

The quality function for this case is displayed at the upper panel 
of Figure~\ref{seismo}.

\subsection{Case~(2): mixed $\ell= 1$ and $\ell=  2$ modes}

In this case we consider that the observed periods are associated with
a mix of  $\ell= 1$ and $\ell=  2$ modes.  We perform a  period fit in
which the value of $\ell$ for  the theoretical periods is not fixed, 
but  instead is  obtained as  a result  of our  period fit
procedure, with  allowed  values of $\ell=  1$ and $\ell=  2$. The
solution is displayed in  the central panel of Figure~\ref{seismo} and
has  $M_*= 0.556\,M_\odot$,  $T_\mathrm{eff}=  128\,300$~K, $L/L_\odot$=
156. The agreement between theoretical  and observed modes is shown in
Table \ref{table-case2}.

\begin{table}
\centering
\begin{tabular}{cccc}
$\Pi_{\rm obs}$& $\Pi_{\rm theor}$&  $\ell$ & $k$ \cr
439.2&439.2 &1& 18\cr
457.2&457.5 &2& 34\cr
523.5&523.3 &1& 22\cr
\end{tabular}
\caption{Observed and theoretical periods for the best model
fit in Case~2.}
\label{table-case2}
\end{table}

\subsection{Case~(3): all $\ell=  2$ modes}
\begin{figure}
\centering
\includegraphics[width=0.5\textwidth]{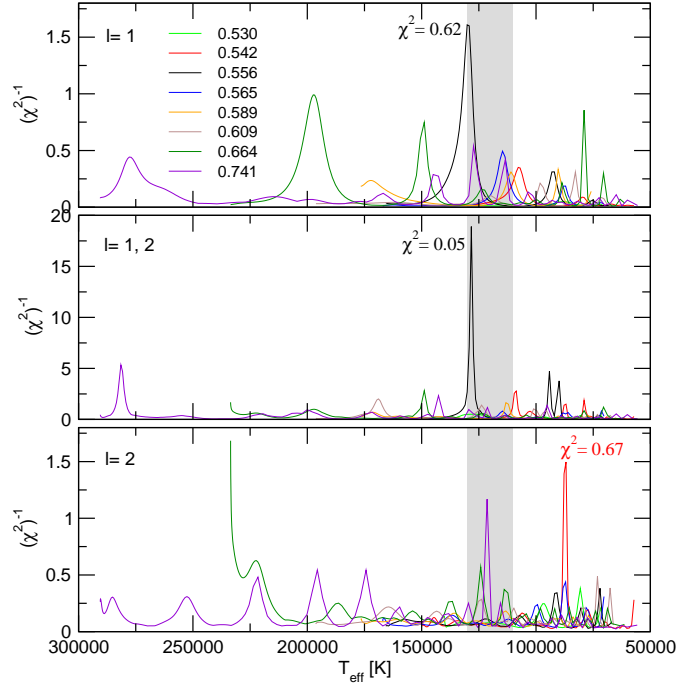}
\caption{The inverse  of the  quality function of  the period  fits in
  terms  of  the  effective   temperature.  The  vertical  gray  strip
  indicates the spectroscopic $T_\mathrm{eff}$ and its uncertainties.  Upper panel
  corresponds to the case in which the three periods are associated to
  $\ell= 1$ modes, middle panel  shows the situation in which there is
  a mix of $\ell=1$ and $\ell=  2$ modes, and lower panel displays the
  case in which the three modes are $\ell= 2$.} 
\label{seismo}
\end{figure}

Finally,  we  assume  that   all  three  identified  periods  are
associated  with  $\ell= 2$,  even  though  it  is improbable  that  a
pulsating  pre-white  dwarf  star  shows only  quadrupole  modes.  The
solution is  displayed in the bottom panel  of Figure~\ref{seismo} and
has $M_*= 0.542 M_\odot$, $T_\mathrm{eff}= 86\,900$~K, $L/L_\odot$= 21
(Table \ref{table-case3}).

\begin{table}
\centering
\begin{tabular}{cccc}
$\Pi_\mathrm{obs}$& $\Pi_\mathrm{theor}$&  $\ell$& $k$ \cr
439.2&440.5& 2& 30\cr
457.2&457.1& 2& 31\cr
523.5&524.1& 2& 36\cr
\end{tabular}
\caption{Observed and theoretical periods for the best model
fit in Case~1.}
\label{table-case3}
\end{table}
\noindent This  solution, already unlikely  from the point of  view of
geometrical cancellation,  gives too low of a temperature, compared  with the
spectral  determination. A  secondary solution  is observed  for $M_*=
0.741 M_\odot$ and $T_\mathrm{eff}=  121\,600$ K, but it can be discarded
because of   its   very   high   mass   value,  as   compared   with   the
spectroscopically   inferred  mass  ($0.52   \pm  0.02   M_\odot$)  of
SDSS~J075415.12+085232.18.

The agreement between theoretical and observed periods of the solution
in Case~(2) is excellent, with two  $\ell= 1$ modes and one $\ell= 2$,
and  the  mean  difference  of  0.17~s  is  within  observational  and
theoretical  uncertainties.  This  solution agrees  with  the spectral
temperature,  $T_\mathrm{eff}=  120\,000\pm 10\,000$~K, 
and is within
the  real uncertainty (i.e., including systematic uncertainties) of the spectroscopic
mass: $M_*= 0.52\pm 0.02$.

To estimate the quality of  our best fits,   we compute the
Bayes Information  Criterion [BIC;  \citet{Koen00}]:
\[{\rm BIC}= N_{\rm p} \left(\frac{\log N}{N} \right) + \log \sigma^2\]
where $N_{\rm p}$ is  the number of free parameters, and $N$
the number of  observed periods. The BIC parameter estimates
the absolute quality of the period fit,  by accounting for
situations in which there are different  numbers of observed periods
and free parameters.  In our case,  $N_{\rm p}= 2$ (stellar mass and
effective temperature),  and $N= 3$.  The  smaller  the value of  BIC,
the better the quality  of the fit.  We obtain ${\rm  BIC}= 0.11$ for
Case (1), ${\rm  BIC}= -0.98$ for Case (2),  and ${\rm  BIC}= 0.14$ for
Case (3). The period  fit of Case (2) is excellent, as reflected by the
corresponding BIC value. It could be compared with the BIC value of
current asteroseismological period fits of pulsating white dwarfs
[see, for instance, \citet{Kim11}].

\section{Discussion}

Time  series imaging  show SDSS  J075415.12+085232.18 is  a non-radial
pulsator in  the PG~1159  pre-white dwarf class,  also called  GW Vir.
Its   spectral   effective   temperature   $T_\mathrm{eff}=120\,000\pm
10\,000$~K and C/He=0.33 by number is comparable  to the prototype, and the main
period at  525˜s is also comparable  to the 516˜s  main periodicity of
PG~1159-035.  Its  low pulsation amplitude led to a small number of
periodicities  detected,  contrary to  the  prototype,  which has  the
largest number of independent  pulsations detected after the Sun. 
That PG~1159 stars 
probably have no atmospheric convection layer
might explain
the absence of  combination frequencies, even
when large amplitudes are detected, as in PG~1159--035 itself
\citep{Costa08a}.

 These  stars evolve
fast,  leading  to  substantial  period  change  due  to  cooling  and
contraction, that  should allow  a detectable period  change in  a few
years \citep{Winget83}. Therefore the star should be monitored at least yearly to allow
evolutionary changes determinations \citep{Costa08b}.


\bsp

\label{lastpage}


\begin{thebibliography}{}

\bibitem[\protect\citeauthoryear{Althaus et 
al.}{2005}]{Althaus05} Althaus L.~G., Serenelli A.~M., Panei J.~A., C{\'o}rsico A.~H., Garc{\'{\i}}a-Berro E., Sc{\'o}ccola C.~G., 2005, A\&A, 435, 631 

\bibitem[\protect\citeauthoryear{Althaus et 
al.}{2010}]{Althaus10} Althaus L.~G., C{\'o}rsico A.~H., Isern J., Garc{\'{\i}}a-Berro E., 2010, A\&ARv, 18, 471 

\bibitem[\protect\citeauthoryear{Bischoff-Kim 
\& {\O}stensen}{2011}]{Kim11} Bischoff-Kim A., {\O}stensen R.~H., 2011, ApJ, 742, L16 

\bibitem[\protect\citeauthoryear{Charpinet, Fontaine, 
\& Brassard}{2009}]{Charpinet09} Charpinet S., Fontaine G., Brassard P., 2009, Natur, 461, 501 

\bibitem[\protect\citeauthoryear{Chote et al.}{2014}]{Chote14} 
Chote P., Sullivan D.~J., Brown R., Harrold S.~T., Winget D.~E., Chandler 
D.~W., 2014, MNRAS, 559 

\bibitem[\protect\citeauthoryear{Clemens, Crain, 
\& Anderson}{2004}]{Clemens04} Clemens J.~C., Crain J.~A., Anderson R., 2004, SPIE, 5492, 331 

\bibitem[\protect\citeauthoryear{C{\'o}rsico 
\& Althaus}{2006a}]{Corsico06a} C{\'o}rsico A.~H., Althaus L.~G., 2006, A\&A, 454, 863 

\bibitem[\protect\citeauthoryear{C{\'o}rsico, Althaus, 
\& Miller Bertolami}{2006b}]{Corsico06b} C{\'o}rsico A.~H., Althaus L.~G., Miller Bertolami M.~M., 2006, A\&A, 458, 259 

\bibitem[\protect\citeauthoryear{C{\'o}rsico et 
al.}{2009}]{Corsico09} C{\'o}rsico A.~H., Althaus L.~G., Miller Bertolami M.~M., Garc{\'{\i}}a-Berro E., 2009, A\&A, 499, 257 

\bibitem[\protect\citeauthoryear{C{\'o}rsico et 
al.}{2008}]{Corsico08} C{\'o}rsico A.~H., Althaus L.~G., Kepler S.~O., Costa J.~E.~S., Miller Bertolami M.~M., 2008, A\&A, 478, 869 

\bibitem[\protect\citeauthoryear{C{\'o}rsico et 
al.}{2007a}]{Corsico07a} C{\'o}rsico A.~H., Althaus L.~G., Miller Bertolami M.~M., Werner K., 2007, A\&A, 461, 1095 

\bibitem[\protect\citeauthoryear{C{\'o}rsico et 
al.}{2007b}]{Corsico07b} C{\'o}rsico A.~H., Miller Bertolami M.~M., Althaus L.~G., Vauclair G., Werner K., 2007, A\&A, 475, 619 

\bibitem[\protect\citeauthoryear{C{\'o}rsico et 
al.}{2011}]{Corsico11} C{\'o}rsico A.~H., Althaus L.~G., Kawaler 
S.~D., Miller Bertolami M.~M., Garc{\'{\i}}a-Berro E., Kepler S.~O., 2011, 
MNRAS, 418, 2519 

\bibitem[\protect\citeauthoryear{Costa et 
al.}{2008a}]{Costa08a} Costa J.~E.~S., et al., 2008, A\&A, 477, 627 

\bibitem[\protect\citeauthoryear{Costa 
\& Kepler}{2008b}]{Costa08b} Costa J.~E.~S., Kepler S.~O., 2008, A\&A, 489, 1225 

\bibitem[\protect\citeauthoryear{Kanaan, Kepler, 
\& Winget}{2002}]{Kanaan02} Kanaan A., Kepler S.~O., Winget D.~E., 2002, A\&A, 389, 896 

\bibitem[\protect\citeauthoryear{Kawaler 
\& Bradley}{1994}]{Kawaler94} Kawaler S.~D., Bradley P.~A., 1994, ApJ, 427, 415 

\bibitem[\protect\citeauthoryear{Kleinman et 
al.}{2013}]{Kleinman13} Kleinman S.~J., et al., 2013, ApJS, 204, 5 

\bibitem[\protect\citeauthoryear{Koen 
\& Laney}{2000}]{Koen00} Koen C., Laney D., 2000, MNRAS, 311, 636 

\bibitem[\protect\citeauthoryear{Liebert et 
al.}{1989}]{Liebert89} Liebert J., Wesemael F., Husfeld D., 
Wehrse R., Starrfield S.~G., Sion E.~M., 1989, AJ, 97, 1440 

\bibitem[\protect\citeauthoryear{McGraw et al.}{1979}]{Mcgraw79} 
McGraw J.~T., Liebert J., Starrfield S.~G., Green R., 1979, wdvd.coll, 377 
%
\bibitem[\protect\citeauthoryear{Metcalfe}{2003}]{Metcalfe03} 
Metcalfe T.~S., 2003, BaltA, 12, 247 

\bibitem[\protect\citeauthoryear{Miller Bertolami 
\& Althaus}{2006}]{Bertolami06} Miller Bertolami M.~M., Althaus L.~G., 2006, A\&A, 454, 845 

\bibitem[\protect\citeauthoryear{Quirion}{2009a}]{Quirion09a} 
Quirion P.-O., 2009, CoAst, 159, 99 

\bibitem[\protect\citeauthoryear{Quirion, Fontaine, 
\& Brassard}{2009b}]{Quirion09b} Quirion P.-O., Fontaine G., Brassard P., 2009, JPhCS, 172, 012077 

\bibitem[\protect\citeauthoryear{Schwarz et 
al.}{2004}]{Schwarz04} Schwarz H.~E., et al., 2004, SPIE, 5492, 
564 

\bibitem[\protect\citeauthoryear{Stetson}{1991}]{Stetson91}
Stetson P.~B., 1991, ESOC, 38, 187 

\bibitem[\protect\citeauthoryear{Thompson \& Mullally}{2009}]{Thompson09}
Thompson S.~E., Mullally F., 2009, Journal of Physics: Conference Series, 172, 012081 

\bibitem[\protect\citeauthoryear{Valdes}{1998}]{Valdes98} Valdes 
F.~G., 1998, ASPC, 145, 53 

\bibitem[\protect\citeauthoryear{Valdes 
\& Tody}{1998}]{ValdesTody98} Valdes F.~G., Tody D., 1998, SPIE, 3355, 497 

\bibitem[\protect\citeauthoryear{Werner, Heber, 
\& Hunger}{1989}]{Werner89} Werner K., Heber U., Hunger K., 1989, LNP, 328, 194 

\bibitem[\protect\citeauthoryear{Werner, Rauch \& Kepler}
{2014}]{Werner14} Werner K., Rauch T., \& Kepler S.O., A\&A, 564, A53

\bibitem[\protect\citeauthoryear{Winget 
\& Kepler}{2008}]{Winget08} Winget D.~E., Kepler S.~O., 2008, ARA\&A, 46, 157 

\bibitem[\protect\citeauthoryear{Winget et al.}{1991}]{Winget91} 
Winget D.~E., et al., 1991, ApJ, 378, 326 

\bibitem[\protect\citeauthoryear{Winget, Hansen, 
\& van Horn}{1983}]{Winget83} Winget D.~E., Hansen C.~J., van Horn H.~M., 1983, Natur, 303, 781 

\bibitem[\protect\citeauthoryear{Woudt, Warner, 
\& Zietsman}{2012}]{Woudt12} Woudt P.~A., Warner B., Zietsman E., 2012, MNRAS, 426, 2137 

\end{thebibliography}
\end{document}